\documentstyle[prl,aps,multicol,epsf]{revtex}
\begin{document} 
\draft

\title{ Tunneling into Ferromagnetic Quantum Hall States:
Observation of a Spin Bottleneck}

\author{H. B. Chan and R. C. Ashoori}
\address{Department of Physics, Massachusetts Institute of
Technology, Cambridge, Massachusetts 02139}
\author{L. N. Pfeiffer and K. W. West}
\address{Bell Laboratories, Lucent Technologies, 
Murray Hill, New Jersey 07974 }
\maketitle
\date{Received}

\begin{abstract}
We explore the characteristics of equilibrium tunneling 
of electrons from a 3D electrode into a high mobility
2D electron system. For most 2D Landau level filling 
factors, we find that tunneling can be characterized by 
a single, well-defined tunneling rate. However, for 
spin-polarized quantum Hall states ( $\nu$ = 1, 3 and 1/3) 
tunneling occurs at two distinct rates that differ by 
up to 2 orders of magnitude. The dependence of the 
two rates on temperature and tunnel barrier thickness 
suggests that slow in-plane spin relaxation creates a 
bottleneck for tunneling of electrons.
\end{abstract}


%comment out the following 2 lines if single column
\begin{multicols}{2}
\narrowtext

The interplay between Zeeman coupling of electronic spins 
to an applied magnetic field and Coulomb interactions among 
electrons leads to remarkable spin configurations of 
quantum Hall systems. For instance, around quantum 
Hall filling factor $\nu$ = 1, powerful 
exchange interactions align electron spins 
to form a nearly perfect ferromagnet \cite{Ando}.  
Theorists predict that the 
elementary charge excitations of this $\nu$ = 1 
quantum Hall state consist of spin textures known as 
Skyrmions \cite{Sondhi}.  The small value of the 
Zeeman energy compared to the Coulomb energy in 
GaAs gives rise to the appropriate conditions for 
the formation of Skyrmions. Nuclear spin 
resonance and magneto-optical absorption 
experiments \cite{Barrett} have shown that the 
spin polarization of the 2D electrons attains 
a maximum at $\nu$ = 1 and falls off sharply 
on either side. This rapid 
loss of spin polarization away from $\nu$ = 1 provides 
the strongest evidence for the existence of 
Skyrmions. Transport and heat capacity
measurements \cite{transport}
offer additional support for the Skyrmion picture. 

Tunneling experiments have demonstrated a capability 
to probe electron-electron 
interactions. For instance, tunneling of electrons into 
2D systems in a magnetic field displays 
characteristics of a pseudogap 
\cite{Ashoori,Dolgo,Chan,Eisenstein} created by 
Coulomb interactions among electrons. 
Given the measured and predicted richness of the spin properties 
of quantum Hall systems, we decided to explore whether 
tunneling could also prove useful for revealing effects of electronic 
spins \cite{Fertig}. Such study should prove 
most interesting for the ferromagnetic quantum Hall states, 
but experimental data for tunneling in 
these regimes have been limited. The major obstacle is 
that the in-plane conductance of the 2D 
system drops to near zero around $\nu$ = 1. As a result 
the tunneling charge cannot be collected 
and measured via conduction in the 2D plane. It is possible 
to use capacitance techniques to 
circumvent this problem \cite{Ashoori,Dolgo}.  However, 
complete characterization involves time-resolved 
measurements described here or measurements over 
broad frequency range that have not been 
previously performed on high mobility samples.

In this letter, we describe measurements of tunneling 
from a 3D electrode into a high mobility 2D electron 
system in a GaAs/AlGaAs heterostructure at $\nu$ = 1. 
Using a novel capacitance technique reported 
previously \cite{Chan}, we detect the tunneling 
current into both localized and delocalized states. 
Here, we focus on the effects of electronic spins on 
tunneling by detecting the equilibrium tunneling of 
electrons in real time, instead of studying the 
tunneling pseudogap through conventional measurement 
of non-linear I-V curves. We observe that the process 
of electron tunneling into ferromagnetic quantum Hall 
states differs qualitatively from tunneling into other 
filling fractions: electrons tunnel into ferromagnetic 
quantum Hall states at two distinct rates. Some electrons 
tunnel into the 2D system at a fast rate while the rest 
tunnel at a rate up to 2 orders of magnitude slower. We 
observe such novel double-rate tunneling only in spin 
polarized quantum Hall states ($\nu$ = 1, 3 and $\leq$1/3) 
in samples of highest mobility. This effect does not 
appear at even-integer filling fractions. Our detailed 
study of the dependence of the two rates on 
temperature, magnetic field and tunnel barrier 
thickness indicates that slow in-plane spin 
relaxation leads to a bottleneck for tunneling and 
gives rise to the double tunneling rate 
phenomenon.

Figure 1a shows a schematic of our samples. The 
following sequence of layers is grown 
on n+ GaAs substrate: 6000 $\AA$ n+ GaAs, 300 $\AA$ 
GaAs spacer layer, AlGaAs/GaAs tunnel barrier, 
175 $\AA$ GaAs quantum well, 700 $\AA$ AlGaAs 
(undoped) blocking barrier and 1.3 $\mu$m n+ GaAs 
cap layer. Samples A and C have 
AlGaAs/GaAs superlattice tunnel barriers of thickness 
193 $\AA$ and 147 $\AA$ respectively. 
For sample B, the tunnel barrier is made 
of 130 $\AA$ AlGaAs. A major 
advantage of our structure is the complete absence of 
silicon dopants in the AlGaAs layers, 
eliminating the main source of disorder in the 2DEG. 
Electrons are attracted into the quantum 
well from the bottom n+ GaAs electrode by application 
of a positive dc bias to the cap layer. As 
a result, the mobility of our  
samples is expected to 
be higher than $10^{6} cm^{2}V^{-1}s^{-1}$,
\begin{figure}
%comment out the next two lines if don't want to incorporate figure
\epsfxsize=\linewidth
\epsfbox{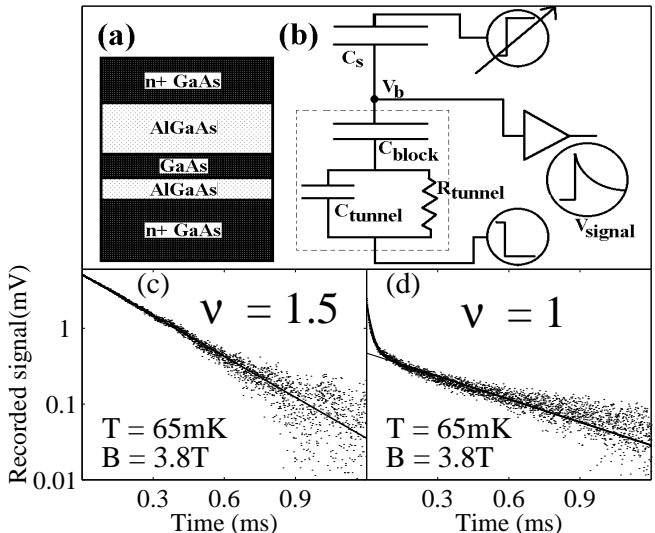}
\caption{(a) Structure of our samples. (b) External circuit 
used to measure $R_{tunnel}$. The sample can be modeled 
by linear circuit elements (box) when the excitation voltage is 
smaller than {\it kT}. (c) Recorded signal (amplification 
of $V_{b}$) decay exponentially at $\nu$ = 1.5. The line is an 
exponential fit to the data. (d) Recorded signal is non-exponential 
at $\nu$  = 1. The thin line is an exponential fit to the data. The 
thick line is a fit to the data using Eq. (1).}
\end{figure}
%if single column, remove the following noindent command
\noindent which is 
consistently achieved in modulation doped GaAs/AlGaAs 
quantum well heterostructures grown in the same MBE 
machine. The dc 
bias to the cap layer also permits variation of the 2D 
electron density from depletion to 
$3 \times 10^{11} cm^{-2}$.

An earlier experiment \cite{Chan} measured the single 
particle density of states of a similar 
structure with lower 2D electron mobility in a magnetic 
field using ``time domain capacitance spectroscopy.''  We 
use the same technique to study the high mobility samples. 
Here, we focus on``zero-bias'' tunneling into the 2DEG 
measured by applying excitation voltages smaller than 
{\it kT}. In this equilibrium tunneling 
regime \cite{Ashoori}, 
we model the tunnel barrier by a capacitor $C_{tunnel}$ 
shunted by a resistor $R_{tunnel}$, while a capacitor 
$C_{block}$ represents the blocking barrier (Fig. 1b). 
Figure 1b also shows the capacitance bridge used to 
measure $R_{tunnel}$. Voltage steps of opposite 
polarity are applied to the top electrode of the sample 
and to one plate of a standard capacitor $C_{s}$. 
The other plate of $C_{s}$ and the bottom electrode of 
the sample are electrically connected, and the 
voltage $V_{b}$ at this balance point is amplified and 
recorded as a function of time. When the 
excitation voltage amplitude is smaller than 
{\it kT}, the 
tunneling resistance $R_{tunnel}$ is independent of 
voltage across the tunnel barrier. The equivalent circuit 
of the bridge consists of linear circuit 
elements and therefore we expect $V_{b}$ to decay 
exponentially. 

Figure 1c plots on a semi-log scale the recorded voltage 
as a function of time at $\nu$ = 1.5. 
%and a field of 3.8 T. 
The signal decays exponentially 
for more than 2 orders of magnitude. In 
general, we observe such agreement with an exponential 
decay when $\nu$ is close to half integer. 
This indicates that for filling factors at which the 2DEG 
is compressible, electrons tunnel into the 
2DEG at a single rate and the equivalent circuit model 
in Fig. 1b adequately describes the 
sample. Figure 1d shows a drastically different recorded 
signal at $\nu$ = 1. The decay is clearly 
non-exponential. We can fit it well with a sum of two 
exponential decays with different time 
constants and prefactors:
\begin{displaymath}
V(t) = A_{1} exp(-t/\tau_{1}) + A_{2} exp(-t/\tau_{2}) \		
$ (1)$
\end{displaymath}
In other words, at $\nu$ = 1 electrons tunnel from the 3D
electrode into the 2DEG at two distinct 
rates. Some electrons tunnel at a fast rate while the rest 
tunnel at a significantly slower rate. We 
emphasize that the measurement is performed in the linear 
response limit of $R_{tunnel}$ by applying 
excitation voltage across the tunnel barrier (8.9 $\mu$V) 
comparable to the temperature (65 mK). This 
eliminates the possibility that the non-exponential relaxation 
at $\nu$ = 1 is due to a voltage 
dependent $R_{tunnel}$ caused by the magnetic field induced 
energy gap in tunneling \cite{Ashoori,Dolgo,Chan,Eisenstein}.

Figure 2 shows the dependence of relaxation rates on gate 
voltage at a fixed magnetic 
field of 3.8 T. At each gate voltage in Fig. 2, we record 
a time trace similar to the ones in Fig. 1c 
and 1d. For gate voltages at which we can fit the time 
trace by a single exponential decay as in 
Fig. 1c, we plot the relaxation rate as a hollow square. 
When it is necessary to use a sum of two 
exponential decays (Eq. (1)) to fit the signal as 
in Fig. 1d, filled triangles and circles 
represent the corresponding fast and slow relaxation 
rates (1/$\tau_{1}$ and 1/$\tau_{2}$) obtained respectively. 
Figure 2 indicates that tunneling occurs at two distinct 
rates near integer Landau level filling factors, 
while electrons tunnel at a single rate 
when the 2DEG is compressible near half 
integer fillings. 

At integer filling factors, the in-plane 
conductance vanishes as the 
electronic states at the chemical potential become 
localized. Inhomogeneity, such as monolayer 
fluctuations in the tunnel barrier thickness, results 
in non-uniform tunneling rates into different 
lateral positions of the 2D plane. In Fig. 2, the 
two relaxation rates at $\nu$ = 2 differ approximately 
by a factor of three and can be explained well by this 
argument. In contrast, the fast and slow 
relaxation rates at $\nu$ = 1 differ by about a factor 
of 60. Relaxation rate differences of such 
magnitude cannot be explained by fluctuations in the 
tunnel barrier thickness. Moreover, the ratio between the two 
relaxation rates also behaves differently around $\nu$ =1 
and $\nu$ = 2 as $\nu$ deviates from exact integer 
value. In Fig. 2, the ratio of the two rates remains 
almost constant around $\nu$ = 2. On the other 
hand, this ratio increases as $\nu$ approaches 1, 
attaining a peak value of 60 at $\nu$ = 1. The inset of Fig. 2 
illustrates the difference between a time trace at 
$\nu$ = 1 and $\nu$ = 2. Both traces decay at a comparable rate
initially (with time constants 
$\sim$ 10 $\mu$Sec), whereas only 
\begin{figure}
%comment out the next two lines if don't want to incorporate figure
\epsfxsize=\linewidth
\epsfbox{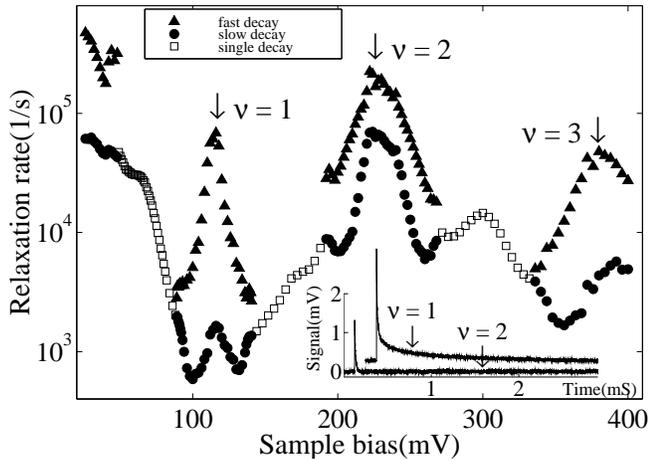}
\caption{Dependence of the relaxation rate of the exponential decay 
on sample bias for sample B at 3.8 T and 65 mK. 
Inset: Comparison of recorded signal at $\nu$ =1 and $\nu$ = 2.} 
\end{figure}
%if single column, remove the following noindent command 	
\noindent the $\nu$ = 1 
signal contains an additional slower decaying 
component with a time constant of about 600 $\mu$Sec. 

The $\nu$ = 1 and $\nu$ = 2 quantum Hall states have 
the common characteristic that an energy 
gap exists at the chemical potential, albeit of 
different origins. At $\nu$ = 2, the cyclotron gap is 
present even when correlation effects are 
neglected. On the other hand, the existence of 
an energy gap at $\nu$ = 1 is a many body phenomenon. 
The interactions among electrons lead to 
ferromagnetic order and the formation of an exchange 
energy gap. In our experiment, we 
measure equilibrium tunneling by applying excitation 
voltages at least 100 times 
smaller than the Coulomb energy. In an ideal 2D system 
without disorder at $\nu$ = 1, there are no 
states at the chemical potential into which electrons 
can tunnel. Any tunneling current detected must arise 
from broadening of the Landau levels due to disorder. 
Consider a 2D system with 
inhomogeneous density. When the bulk filling factor is one, 
regions with local density higher 
(lower) than the bulk density have filling fraction $\nu >$ 1 
($\nu <$ 1) into which electrons with minority 
(majority) spin tunnel. To our knowledge, theories do not
presently predict that the tunneling rates of 
electrons with spin up and down are significantly different. 
While a difference in the tunneling 
rates for electrons with opposite spins can lead to 
observation of two relaxation rates in our  	
experiment, we show below that this hypothesis is 
inadequate to explain our data. 

Figure 3a plots the relaxation rate as a function of 
gate voltage at 5.7 T. Similar to the 
data at a lower field in Fig. 2, tunneling occurs at 
two distinct rates around $\nu$ = 1. In addition to 
the relaxation rates, we also show the prefactors of 
the exponential fits ($A_{1}$ and $A_{2}$ in Eq. 
(1) scaled by a constant factor) in Fig. 3b. 
Around $\nu$ = 1, $A_{1}$ and $A_{2}$ are proportional to the 
amount of charge tunneling at the fast and slow rates 
respectively. For the slow decay, the 
prefactor (plotted as circles) has a minimum 
at $\nu$ = 1 while the prefactor for the fast decay (plotted
\begin{figure}
%comment out the next two lines if don't want to incorporate figure
\epsfxsize=\linewidth
\epsfbox{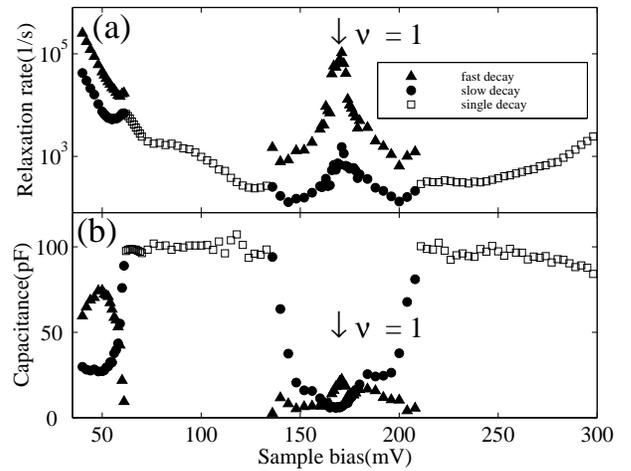}
\caption{(a) Relaxation rate vs. sample bias for sample B at 5.7 T. 
(b) Sample capacitance charged at the fast and slow 
rates vs. sample bias. The two capacitance contributions refer to 
the ratio of the charge tunneling at the fast and slow rates 
(proportional to $A_{1}$ and $A_{2}$ respectively in Eq. (1)) 
to the constant excitation voltage (9 $\mu$V).}
\end{figure}
%if single column, remove the following noindent command
\noindent as triangles) instead has a maximum. 
Consider a 2D system with inhomogeneous density 
at bulk filling factor $\nu$ = 1. As the bulk density 
is increased the fraction of regions with local 
filling factor $\nu <$ 1 decreases monotonically and 
vice versa for regions with local $\nu >$ 1. If 
electrons with majority and minority spins tunnel at 
different rates, we expect the prefactors of 
the fast (slow) decay to be an increasing (decreasing) 
function of bulk density around $\nu$ = 1, in 
contrary to Fig. 3b. Therefore the observation of two 
relaxation rates at $\nu$ = 1 cannot be trivially 
explained by a difference in the tunneling rates for 
electrons with majority and minority spins.

Figure 4 shows the temperature dependence of the 
two relaxation rates at $\nu$ = 1 for three 
magnetic field strengths. At each magnetic field, 
we adjust the density to maintain the filling factor 
at $\nu$ = 1. Both the slow and fast rates 
have rather weak temperature dependence at low 
temperature for all three magnetic fields. The weak 
temperature dependence of the slow 
relaxation rate persists up to a temperature beyond 
which the slow relaxation rate speeds up 
significantly and the double tunneling rate phenomenon 
recedes. This onset of strong 
temperature dependence shifts to a higher temperature 
as the magnetic field is increased. From 
Fig. 4, we identify the characteristic temperature 
$T_{C}$ at which the slow rate rises to a value equal 
to the geometric mean of the two tunneling rates at 
the lowest temperature (as indicated by the 
arrows) and plot it as a function of magnetic field in 
the inset of Fig. 4. In this range of magnetic field, 
$T_{C}$ ($\sim$350 mK at 4.5 T) sets an energy scale 
that is much smaller than the Coulomb energy and the 
cyclotron energy (106 K and 90 K at 4.5 T respectively). 
The only obvious energy scale comparable to $T_{C}$ 
is the Zeeman energy (1.3 K at 4.5 T). In other words 
the development 
\begin{figure}
%comment out the next two lines if don't want to incorporate figure
\epsfxsize=\linewidth
\epsfbox{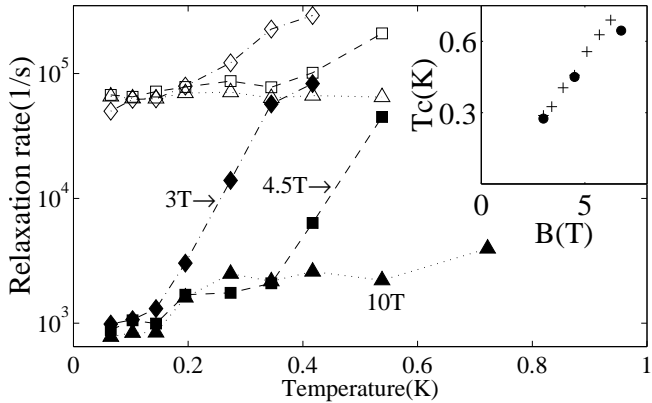}
\caption{Temperature dependence of the fast (hollow) and
 slow (filled) relaxation rates for sample B at $\nu$  = 1 for 3 T
 (diamonds), 4.5 T (squares) and 10 T (triangles). 
 Inset: Characteristic temperature $T_{C}$ (defined in text) 
 vs. magnetic field for sample A (crosses) and sample B (circles).}
\end{figure}
\noindent of the exchange energy 
gap at $\nu$ = 1 is not a sufficient condition for 
tunneling to occur at two rates. For instance, at a 
field of 4.5 T and temperature of 1 K, a minimum in 
the capacitance of the 2DEG is clearly 
observable at $\nu$ = 1, indicating the existence 
of the exchange energy gap. However, as Fig. 4 shows, 
electrons 
no longer tunnel at two rates at this temperature and 
field. This demonstrates that 
spin effects are crucial in explaining 
why tunneling occur at two rates at $\nu$ = 1. 

Possible explanations of the double tunneling rate 
phenomenon at $\nu$ = 1 can generally be 
classified into two approaches. In the first approach, 
electrons are assumed to tunnel into the 2D 
system at a fast rate. The system then undergoes 
certain form of relaxation, possibly spin related, 
within the 2D plane at the slow rate. Through the 
spin relaxation, the 2D system is able to accept 
more electrons tunneling from the 3D electrode giving 
rise to a second, slower tunneling rate. 
Unlike the fast tunneling rate, the slow relaxation 
is expected to have no dependence on the 
thickness of the tunnel barrier. A second approach 
considers the $\nu$ = 1 system bifurcating into 
separate regions into which electrons tunnel at 
different rates. In contrast to the first scenario, 
the ratio of the two tunneling rates should remain 
constant as the tunnel barrier thickness is varied.

In order to differentiate between these two 
possibilities, we measure the relaxation rates 
for samples grown in the same MBE machine with various 
tunnel barrier thickness. 
The results are listed in Table I. 
At $\nu$ = 1/2, we observe a single relaxation rate
in all samples. The relaxation rate increases by 
more than 3 orders of magnitude as the tunnel barrier 
becomes more transparent. In contrast, the slow rate 
at $\nu$ = 1 is relatively insensitive to the 
thickness of the tunnel barrier, varying by less than 
a factor of 10. This provides strong evidence 
that the slow tunneling rate at $\nu$ = 1 is largely
due to relaxation within the 2D plane. 
Since the slow tunneling rate only appears in spin 
polarized quantum Hall states at temperatures 
lower than the Zeeman energy, we describe it as 
arising from a``spin bottleneck'' in which in-plane 
spin relaxation must proceed before additional 
electrons can tunnel into the system.

One example of in-plane relaxation that might 
be relevant is the formation of Skyrmions 
around $\nu$ = 1. For a perfectly uniform system 
precisely in the $\nu$ = 1 ferromagnetic state, 
tunneling injects a single minority spin because 
the thickness of the tunnel barrier ensures that electrons 
tunnel as single entities. Since this is not the 
lowest energy excitation, over time the 2D system 
can lower its energy by flipping more spins to create 
Skyrmions. Because the energy of the 2D 
system is lowered by Skyrmion formation, more 
electrons tunnel from the 3D electrode to keep 
the chemical potentials on the two sides of the 
tunnel barrier aligned. When the time scale for 
spin relaxation is long, the intermediate stage 
forms a bottleneck and temporarily prevents more 
electrons from tunneling. The slow relaxation time 
of $\sim$1 ms is comparable to electron spin 
relaxation times measured in a recent NMR 
experiment \cite{Kuzma}. MacDonald \cite{MacDonald} 
considers spin-up and spin-down electrons tunneling
into the $\nu$ = 1 state with equivalent tunneling 
rates. They must, however, be added to the system 
according to a certain ratio in order to form Skyrmions. 
For instance, creation of a Skyrmion consisting of 3 
flipped spins requires the addition of 4 minority 
spins together with the removal of 3 majority spins. 
This constraint leads to non-equilibrium spin accumulation 
in the tunneling process, and MacDonald predicts a ratio 
of fast and slow relaxation rates in good 
agreement with our data.  

Finally, we note that other researchers \cite{Dolgo} reported tunneling 
relaxation measurements on similar structures around $\nu$ = 1 
and did not observe the bifurcation of rates described here. 
We believe that this experiment was performed over a 
range of frequencies too low and narrow to permit 
detection of the fast rate, and we speculate that their 
data reflect the behavior of the slow relaxation. 
We thank A. H. MacDonald, L. S. Levitov, B. I. Halperin, 
S. V. Iordanski, P. A. Lee and 
X. G. Wen for useful discussions. This work is supported 
by the ONR, JSEP-DAAH04-95-1-0038, the Packard Foundation, 
NSF DMR-9357226 and DMR-9311825.

\begin{table}
\begin{tabular}{|c|c|c|c|} 
Sample & $\nu$ = 1/2 & $\nu$ = 1 slow rate 
& $\nu$ = 1 fast rate \\  \hline
A      &  3.7 (1/s) &   155 (1/s)			    
&      4167 (1/s)	         \\  	
B      & 332 (1/s)         &    848 (1/s)               
&        75060 (1/s)         \\
C      & 6870 (1/s)        &   1380 (1/s)			    
&       out of range	     \\
\end{tabular}
\caption{Relaxation rates at 6.6T for samples with different tunnel 
barrier thickness at $\nu$ = 1/2 and at $\nu$ = 1.} 
\end{table}

\end{multicols}

\end{document}